\documentclass[a4paper, amsfonts, amssymb, amsmath, reprint, showkeys, nofootinbib, twoside,longbibliography, aps]{revtex4-1}
\usepackage[english]{babel}
\usepackage[utf8]{inputenc}
\usepackage{printlen}
\usepackage[detect-all]{siunitx}
\usepackage{comment}
\usepackage[normalem]{ulem} 
\usepackage[colorinlistoftodos, color=green!40, prependcaption]{todonotes}
\usepackage{amsthm}
\usepackage{mathtools}
\usepackage{physics}
\usepackage{xcolor}
\usepackage{graphicx}
\usepackage[left=23mm,right=13mm,top=35mm,columnsep=15pt]{geometry} 
\usepackage{adjustbox}
\usepackage{placeins}
\usepackage[T1]{fontenc}
\usepackage{lipsum}
\usepackage{csquotes}
\usepackage[pdftex, pdftitle={Article}, pdfauthor={Author}]{hyperref} 

\usepackage{soul,xcolor}

\begin{document}



\title{Microresonator and Laser Parameters Definition via Self-Injection Locking}

\author{Artem E. Shitikov\textsuperscript{1,2}}
\email{Shitikov@physics.msu.ru}
\author{Oleg V. Benderov\textsuperscript{3}}
\author{Nikita M. Kondratiev\textsuperscript{1}}
\author{Valery E. Lobanov\textsuperscript{1}}%
\author{Alexander V. Rodin\textsuperscript{3}}
\author{Igor A. Bilenko\textsuperscript{1,2}}
\affiliation{\textsuperscript{1}Russian Quantum Center, 143026 Skolkovo, Russia
}
\affiliation{\textsuperscript{2}Faculty of Physics, Lomonosov Moscow State University, 119991 Moscow, Russia}
\affiliation{\textsuperscript{3}Moscow Institute of Physics and Technology, 141701, Dolgoprudny, Russia}
\date{\today} 

\begin{abstract}


Self-injection locking to a high-quality-factor microresonator is a key component of various up-to-date photonic applications, including compact narrow-linewidth lasers and microcomb sources. For optimal construction of such devices it is necessary to know the parameters of the locking mode, which is a challenging task. We developed and verified experimentally an original technique based on the fundamentals of the self-injection locking effect, which allows determining the key parameters of the mode of the microresonator (quality factor, vertical index) as well as laser diode parameters. We demonstrated that this method can be used in the spectral ranges where conventional methods are not applicable, for example, in the mid-IR. We stabilized a 2.64 $\mu$m laser diode by a high-Q whispering gallery mode microresonator made of crystalline silicon. Using the elaborated technique, the quality factor of the microresonator was determined to be $5\cdot10^8$. 
\end{abstract}
\maketitle

\section{Introduction}

The effect of the self-injection locking (SIL) has been utilized for many years in radiophysics and microwave electronics to stabilize the devices and to increase their spectral purity \cite{Ohta1,Ota1,Chang1,magnetron1, gyrotron1,6999932,7119883}. For more than the last three decades, it has also been studied and actively applied in optics and laser physics \cite{osti_6336783,Lang1980,Belenov_1983,Patzak1983,Agrawal1984,Tkach1986, Dahmani:87,Hollberg:1987,Li:1989,hemmerich90oc,Hemmerich:94, hjelme91jqe}. The most interesting results were demonstrated with whispering gallery mode (WGM) microresonators \cite{BRAGINSKY1989393,PhysRevA.70.051804,1588878,Savchenkov:07,doi:10.1002/lpor.201000025,Lin:14,Henriet:15,strekalov:2016,GRUDININ200633,lecaplain2016mid,Shitikov:18}, combining a large quality factor in a wide spectral range with small size and low environmental sensitivity. Nowadays, this effect is a key ingredient of various up-to-date photonic applications. First, the Rayleigh backscattering in optical microresonators \cite{ Gorodetsky:00} provides a passive frequency-selective fast optical feedback to the laser diode resulting in significant laser phase noise suppression and linewidth reduction. Recent research has demonstrated passive stabilization of single-frequency \cite{VASSILIEV1998305,Vassiliev2003,Liang:10,Liang2015,Dale:16,Savchenkov:19a} or even multifrequency \cite{Donvalkar_2018,Galiev:18,Pavlov2018,Savchenkov:19} semiconductor lasers to subkilohertz linewidths with WGM microresonators in different spectral ranges, from UV to mid-IR. Moreover, it was shown that such stabilized laser diodes can be used as a pump source for the generation of the microresonator-based frequency combs \cite{Pavlov2018}. High attainable Q-factors and small mode volume of the WGM allows decreasing the soliton generation power threshold to several $\mu W$, which opens the way to compact energy-efficient devices. The development of such compact microcomb sources is one of the hottest topics of modern photonics which attracts researchers worldwide \cite{Pavlov2018, raja2019electrically, boust2019compact, briles2019generation, Shen2020, voloshin2019dynamics}. Such devices are of paramount importance in numerous areas of modern science and technology, such as coherent communication \cite{Marin-Palomo2017,fulop2018high}, ultrafast optical ranging and LIDARs \cite{Suh2018,Trocha2018,Riemensberger2020}, high-precision spectroscopy \cite{Suh600,Yang:19}, astrophysics \cite{Obrzud2019,Suh2019}, low-noise microwave synthesis \cite{Liang2015}, and optical clocks \cite{Papp:14,Newman:19}.

Constructing such microresonator-based devices operating in the SIL regime, one may face a challenging problem of the accurate determination of the locking microresonator mode parameters. Its quality factor is a crucial parameter that determines the effectiveness of the SIL and overall device performance, since it determines the resulting linewidth of the laser source defining the stabilization coefficient \cite{Kondratiev:17} and the threshold of the nonlinear effects \cite{kippenberg2018dissipative}. The highest Q-factor can be achieved in WGM crystalline resonators (up to 10$^{11}$ \cite{Savchenkov:07}) and application of prism couplers made devices with such resonators perfectly working and robust. There are several well-known ways to measure the quality factor: the full width at half maximum (FWHM) of the resonance curve measurement, the ringdown method based on the recording of free oscillations after pulsed excitation, and evaluation from the linewidth dependence on the microresonator loading \cite{Coupling18}. However, in the SIL regime, such methods are mostly non-applicable. First, they require an optical isolator between the laser and the microresonator, which leads to the need for the disassembly and subsequent assembly of the device or the setup. Second, a determination of the particular WGM that was used for the SIL can be a challenging task in the overmoded resonators. Also, in some spectral ranges, the use of these methods requires the availability of hard-to-find and expensive instruments. 

\begin{figure*}[ht]
\centering
\includegraphics[width=\linewidth]{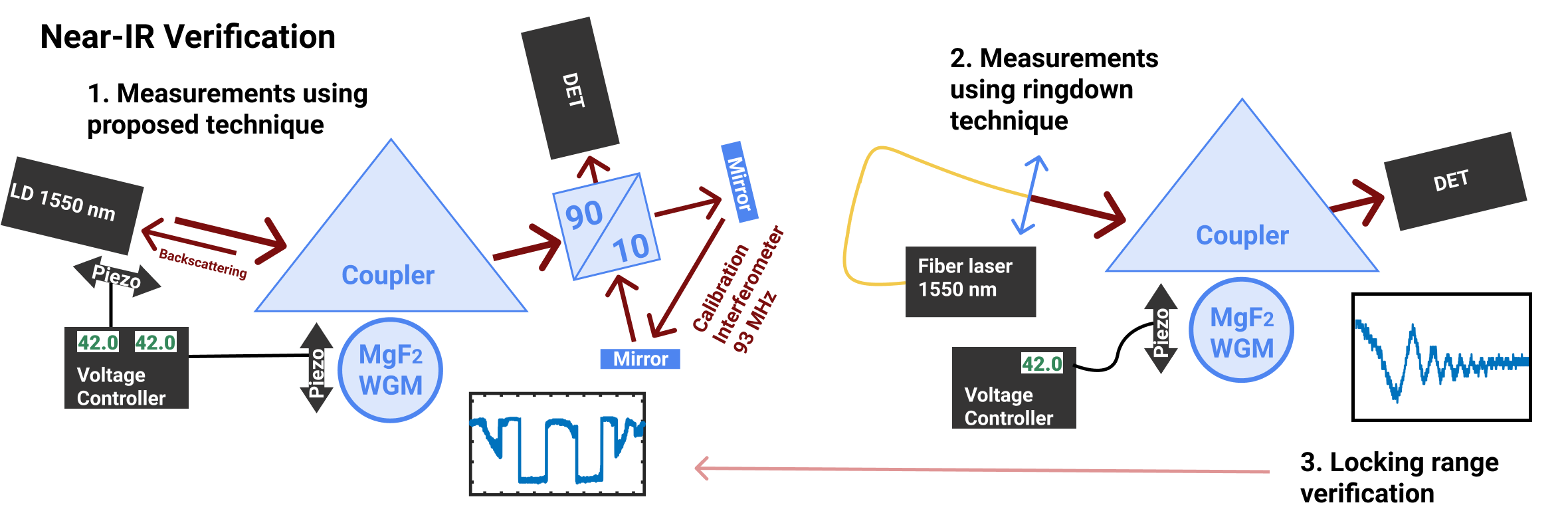}
\caption{Experimental setups in the near-IR. First, we measured the dependence of the resonance width on the gap value between the microresonator and the coupler. Second, we measured the Q-factor of that mode by the ringdown method. Finally, we verified that Q-factor conserves during measurements by measuring the resonance width another time.
}
\label{fig:theBegining}
\end{figure*}

In our work, we analyzed in detail the manifestations of the SIL for different types and parameters of the microresonator modes and showed that SIL can be used as an effective diagnostic tool. We found out that SIL transmission curves depend on the microresonator Q-factor in a fashion that allows extracting it directly from the locking range measurement, which makes it unnecessary to measure microresonator quality factor from the frequency or time response of the unlocked resonator. We developed a new, original method based on the measurement of the SIL bandwidth as a function of the gap between the coupling prism and the microresonator. The possibility of varying the coupling rate by changing the size of the gap gives a useful degree of freedom (inaccessible in the case of Fabry-Perot cavities). The technique also offers a unique opportunity for determining the intrinsic quality factor of the exact WGM used in the SIL regime. Besides, the developed method allows determining the vertical index of the selected mode and provides information about a laser diode parameter -- the output beam coupling rate --  which is of particular importance for accurate SIL modeling and microresonator-based device construction. This technique is based on the fundamentals of SIL and is valid for any wavelength and microresonator and was verified at 1550 nm for MgF$_2$ by the conventional well-adopted ringdown method.
We also compared the elaborated technique with the alternative techniques and showed that it is especially useful for the WGM  Q-factor determination in specific wavelength ranges where it is difficult to do using conventional methods due to the lack of necessary devices (high-speed detectors, narrow-linewidth lasers).

Namely, we successfully applied the technique for silicon resonators in the mid-IR band, avoiding the need to use an expensive narrow-linewidth tunable laser. We demonstrated the first to our knowledge SIL of a laser diode to a WGM microresonator made of crystalline silicon at $2.6\, \mu m$ wavelength, and characterized the effect using the above-mentioned technique. The Q-factor of the WGM was up to $0.5 \cdot 10^9$, and the locking width exceeded $0.6$ GHz.
The mid-IR laser diodes' significant stabilization seems to be very interesting because there is a shortage of affordable, narrow-linewidth laser sources in the enormous range from 2$\ \mu$m to 20$\ \mu$m. Commonly used optical parametric oscillators and quantum cascaded lasers with distributed feedback are rather expensive and hard-to-obtain products. Sub-kHz lasers have great potential in the mid-IR range since it covers the molecular fingerprint region important for high-resolution gas spectroscopy \cite{tokunaga2013probing}, biosensing  \cite{Xia:19}, and even fundamental constants measurements \cite{mejri2015measuring}. It is also promising for free-space optical communication due to better performance than the near-IR light in fog conditions because it suffers less from atmospheric turbulence \cite{hao2017mid}. For optical communication, a narrow-linewidth laser source allows high order modulation to increase the bit-rate \cite{Al-Taiy:14}. 
The demonstration of the high-quality factor in silicon microresonator in the mid-IR is also a new step in silicon photonics. The high nonlinear refractive index, $n_2 = 4.5 \cdot 10^{-18}$ m$^2$/W \cite{bristow2007two_photon_abs}, combined with the absence of the multiphonon absorption up to 8 $\mu$m \cite{hass1977residual} and two-photon absorption from 2.3 $\mu$m \cite{bristow2007two_photon_abs} opens up a new perspective for silicon in the mid-IR since the quality factor of the commonly used fluoride WGM reference cavities can be limited by the multiphonon absorption \cite{lecaplain2016mid}.

\section {SIL-based measurement technique}

\begin{figure*}[ht!]
\centering
\includegraphics[width=\linewidth]{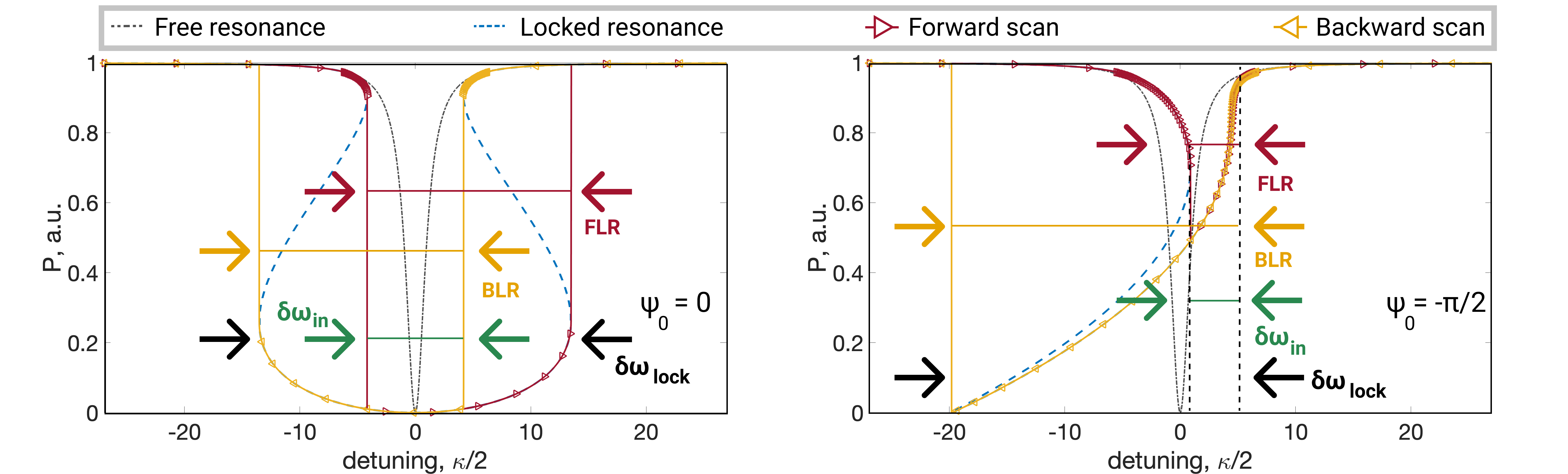}
\caption{Calculated normalized transmitted power dependence on the detuning of the laser frequency from the WGM frequency for the different locking  phases. The sum of the forward and backward locking ranges (FLR and BLR) was measured in experiments. It is a good approximation of the total locking range $\delta\omega_{\rm lock}$ in the case of high Q-factor, when the inner overlap band $\delta\omega_{\rm in}$ is negligible.
}
\label{fig:Detuning}
\end{figure*}

We consider a typical microresonator experimental setup (see Fig. \ref{fig:theBegining}) with piezo elements to control the microresonator coupling and the laser-microresonator distance.
We studied different types of dependencies of the transmitted light intensity from the laser diode current (LI-curves), observed in experiments with a photodiode, which corresponds to a frequency scan of a laser diode. In case of the SIL the LI-curve shape varies with the phase of the backscattering wave $\psi_0$ gained while travelling from the laser to the microresonator and back. 
The locking phase $\psi_0$ strongly affects  the laser stabilization and can be controlled by adjustment the distance between the laser diode and WGM microresonator. In our experiments the locking phase $\psi_0$ was precisely controlled by a piezo element (see Fig. \ref{fig:theBegining}). Calculated dependencies of the transmitted power on the laser detuning from WGM frequency are shown in Fig. \ref{fig:Detuning}. A free resonance is observed in the absence of backscattering (black dotted line in Fig. \ref{fig:Detuning}) and a locked resonance appears when the backscattering causes SIL (blue dashed line in Fig. \ref{fig:Detuning}). In the self-injection locking regime, one may determine the locking range, that is the bandwidth on the LI-curve where the frequency change is suppressed by SIL. At locking phases $\psi_0\in[-\pi/2;\pi/2]$, the locking range is bounded by sharp edges. For non-even locking phases ($\psi_0\neq \pi n$), the shape of the LI-curves during the forward and the backward frequency scans are different, which can be seen in the experiment. This means that we need both scans to capture the whole locking band correctly. The theoretical predictions are shown in Fig. \ref{fig:Detuning} with solid lines with triangle markers. The sum of the forward locking range (FLR), measured at forward scan, and backward locking range (BLR), measured at backward scan, is connected with the locking range $\delta\omega_{\rm lock}$ as ${\rm FLR}+{\rm BLR}=\delta\omega_{\rm lock}+\delta\omega_{\rm in}$ (see Fig. \ref{fig:Detuning}). The FLR-BLR overlap or the inner band $\delta\omega_{\rm in}$ is significant only for low intrinsic quality factor $Q_{int}$ or for overcoupled microresonator.

We note that in the locked state the width of the locking range can be naturally defined using the sharp jumps of the curve. The self-injection locking theory predicts that this width depends on the loaded quality factor of the locking WGM $Q_m$ ($1/Q_m=1/Q_{\rm int}+1/Q_{\rm coupl}$, where $Q_{\rm coupl}$ represents losses added by the coupler that extracts light from resonator) \cite{Kondratiev:17}. Using this theory, we propose a technique, based on measuring of ${\rm FLR}+{\rm BLR}$ dependence on a gap between the cavity and the coupler, which provides information on the locking mode's intrinsic quality factor $Q_{\rm int}$ and its vertical index. Knowing $Q_{\rm int}$ one may estimate $Q_m$ for every gap value. 

Using the expression for the stationary tuning curve for the zero locking phase $\psi_0=0$ \cite{Kondratiev:17}, one may obtain the following expressions for the total locking range width $\delta\omega_{\rm lock}$ and for the width of the inner band $\delta\omega_{\rm in}$ in the case of zero locking phase:   

\begin{align}
\label{Lock}
\delta\omega_{\rm lock} &\approx 3\sqrt{3}\gamma\frac{\kappa_{mc}}{\kappa_m^2}\bar{\kappa}_{do}+\frac{\sqrt{3}}{3}\kappa_m,\\
\label{Inner}
\delta\omega_{\rm in}&\approx 8\sqrt[4]{\frac{\kappa_{mc}\kappa_m\gamma\bar\kappa_{do}}{27}}.
\end{align}
Here, $\gamma$ is the forward-backward wave coupling rate, $\bar{\kappa}_{do}$ is the laser's effective (hot) output beam coupling rate and $\kappa_m = \kappa_{mc} + \kappa_{mi}$ is the microresonator’s mode decay rate, where $\kappa_{mi}$ is determined by the intrinsic losses and $\kappa_{mc}$ is determined by the coupling losses. The corresponding values of the loaded and intrinsic quality factor are defined as $Q_m = \omega/\kappa_m$, $Q_{int} = \omega/\kappa_{mi}$. One can load a microresonator by approaching it to the coupler. The microresonator is called overcoupled when $\kappa_{\rm mc}$ is larger than $\kappa_{\rm mi}$ and undercoupled when $\kappa_{\rm mi}$ is larger than $\kappa_{\rm mc}$. In case of equality of $\kappa_{\rm mi}$ and $\kappa_{\rm mc}$ the coupling is called critical. It should be noted that the expressions \eqref{Lock} and \eqref{Inner} provide good estimations for the widths of the discussed bands for the locking phases $\psi_0\in[-\pi/2;\pi/2]$. In an experiment the locking phase close to zero is the most desirable due to the best performance \cite{Galiev20}. The second term in \eqref{Lock} becomes a significant part of the locking range only in close proximity of the coupler, where $\kappa_{mc}$ is large. However, the locking range \eqref{Lock} is actually determined at the level higher than 1/2 of maximum intensity and, thus, does not correspond to the full width at half-maximum (FWHM) which is usually used for the Lorentzian-like curve characterization and equals to the resonator loss rate. 
During the loading, the $Q_m$ decreases and at some moment the sharp edges of the locking band are blurred and after that FWHM should be measured to characterize the mode. Using the tuning curve expression from \cite{Kondratiev:17}, one can obtain the following expression for the FWHM in the SIL regime:
\begin{equation}
\delta\omega_{\rm FWHM} = 4\gamma\frac{\kappa_{mc}}{\kappa_m^2}\bar\kappa_{do}+\kappa_m.
\label{SILFWHM}
\end{equation}
It can be seen that $\delta\omega_{\rm FWHM}$ is smaller than the full locking range $\delta\omega_{\rm lock}$ for high values of the intrinsic quality factor and can not be observed as it is hidden behind the LI curve jumps, so \eqref{Lock} should be used to estimate the locking width. If the locking width is reduced, a transition from the ``hard'' (with jumps) to ``soft'' (smooth curve) regime occurs. The transition point is when all the three widths are equal to double unlocked microresonator linewidth: $\delta\omega_{\rm lock}=\delta\omega_{\rm in}=\delta\omega_{\rm FWHM}=2\kappa_m$. After this transition occurs, $\delta\omega_{\rm FWHM}$ becomes bigger than the full locking width and the curve has no jumps anymore (Eq. \eqref{Inner} becomes meaningless), making the usage of $\delta\omega_{\rm FWHM}$ reasonable for the width estimations. When the first term of \eqref{SILFWHM} becomes much smaller than the second one, the $\delta\omega_{\rm FWHM}$ becomes equal to the microresonator total loss rate $\kappa_m$, the curve shape becomes Lorentzian, and SIL is totally switched off.

Analyzing the derivative of \eqref{Lock} with respect to the microresonator coupling rate $\kappa_{mc}$, one may evaluate that its maximum takes place at critical coupling $\kappa_{mi} = \kappa_{mc}$. Knowing $\kappa_{mc}$ at maximum locking range one would also know $\kappa_{mi}$. At the same time, $\kappa_{mc}$ can be evaluated from the value of the gap between the microresonator and the coupler $d$ \cite{gorodetsky1994high}:
\begin{align}
\label{Qfactor}
   \kappa_{mc} \approx \frac{\omega}{2} \left(\frac{n^2-1}{n}k\right)^{-3/2}  \exp{(-2kd\sqrt{n^2-1})}\times& \nonumber\\
 \times
 \begin{cases}
    1/\sqrt{\frac{\pi}{1+\sqrt{n^2-1}}}, p=0
 \\
    1/(\sqrt{2}\pi\sqrt{p}), p > 0
\end{cases}
\end{align}
Here both the refractive indices of the coupler and the microresonator are assumed to be equal to $n$, $\omega$ is the pump frequency, $a$ is the radius of the microresonator, $k=\omega/c$, $p$ is the vertical index of the mode. It is assumed that refractive index of the coupler is approximately equal to the refractive index of the microresonator, which is exactly true in our case of a Si resonator with a silicon hemisphere coupler and quite acceptable for MgF$_2$ with a glass prism. 
Thus, the position of the extremum of the curve describing the dependence of the locking range width (or FLR + BLR in our experiments) on the gap between microresonator and coupling element provides us the information about the locking mode's critical coupling.
The accuracy of the technique depends on the accuracy of the maximum of the locking width position corresponding to the coupler position measurements. Note, that according to \cite{Galiev20} critical coupling provides the optimal laser stabilization in most cases, and thus the obtained information can be also used for the fine-tuning of the SIL-based devices using the same piezo element. 

It is worth noting, that according to \eqref{Qfactor}, the coupling-related linewidth of the microresonator at zero gap in the case of high $Q_{int}$ is mostly determined by the vertical index $p$ of the WGM (see Table \eqref{tab:shape-functions} for 2 mm radius for MgF$_2$ at 1550 nm and 1.25 mm radius for Si at 2639 nm). Thus, knowing $\kappa_m$ at zero gap one can evaluate the vertical index of the mode from \eqref{Qfactor}. The maximum difference between $\kappa_{mc}(d=0)$ is for $p = 0$ and $p = 1$, and becomes lower with increase of $p$.
\begin{table}[htbp]
\centering

\caption{Coupling rate $\kappa_{mc}$ of the WGM at zero gap in MHz}
\begin{tabular}{ccccccc}
\hline
 & p=0 & p=1 & p=2  & p=3 & p=4 & p=5 \\
\hline
$MgF_2$ 1550 nm & 159 & 52 & 37 & 30 & 26 & 23\\
$Si$ 2639 nm & 90 & 14 & 10 & 8 & 7 & 6\\

\hline
\end{tabular}
  \label{tab:shape-functions}
\end{table}

Summarizing, to determine a WGM's parameters in the SIL regime, one needs to measure the dependence of the sum of FLR and BLR on the gap between the coupler and microresonator. Also, it is necessary to monitor the change of the LI-curve shape.  There can be two cases: the first one is when the "hard" SIL regime with sharp LI-curve edges is maintained up to the zero gap; the second case is when the "soft" SIL regime with smooth edges appears in the proximity of the coupler.

The fit process of the measured dependence was carried out by the brute force method. In the first case we start with selecting $Q_{\rm int}$ to match the position of the experimental curve maximum with Eq. \eqref{Lock} and Eq. \eqref{Inner} using Eq. \eqref{Qfactor} for several $p$ values. Then we adjust $\gamma\bar\kappa_{do}$ to fit $\rm BLR+FLR$ at critical coupling and choose the best variant at the zero point over the $p$ indices. 
In the second case the fit process is almost the same but the $p$ index can be estimated from the data at zero gap using Table \ref{tab:shape-functions}. Then we select $Q_{\rm{int}}$ to match the maximum of the experimental curve with Eq. \eqref{Lock} and Eq. \eqref{Inner} using Eq. \eqref{Qfactor}. Then,  we adjust $\gamma\bar\kappa_{do}$ in order to match the locking range value at maximum point. The part of the measured dependence related to the "soft" regime should be fitted with $2\delta\omega_{\rm FWHM}$ from Eqs. \eqref{SILFWHM} and \eqref{Qfactor} using the parameters determined in previous steps, and good coincidence indicates the correct choice of the parameters ($p$ in particular).
The uncertainty of the measurements can be added according to the piezo element voltage controller step.

The only alternative for determining the Q-factor in the SIL regime is a technique based on the resonance depth analysis \cite{Coupling18}. That technique is based on similar principle of the coupling rate exponential fitting \eqref{Qfactor}, but with the resonator power transmittance curve. First of all, the resonance depth measurements are sensitive to the coupling of all light into the detector (which can be challenging in many cases) and different detector noises. Secondly, it is necessary to emphasize that it is vital to determine the vertical index of the mode as described above to calculate Q-factor correctly. Finally, it is very natural to measure the locking range in the SIL regime. It can be made very accurately and is not demanding for the equipment parameters. For a brief estimation of the $Q_{\rm int}$ one doesn't need to measure dozens of points: it is necessary to determine the zero-gap point for the vertical index $p$ determination, and the critical coupling point to relate the coupling to the intrinsic loss. FLR+BLR at these points provides enough information to completely fit with \eqref{Lock}, \eqref{SILFWHM} and \eqref{Qfactor}.

Once the locking range for the given resonator and coupling rate is obtained, it can be used to characterize the laser diode. In the case of the high-Q WGM the second term in \eqref{Lock} is negligible and for the effective (hot) output beam coupling rate $\bar{\kappa}_{do}$ one may obtain from \eqref{Lock}:
\begin{equation}
\bar{\kappa}_{d_o} = \frac{1}{3\sqrt{3}}\frac{\delta\omega_{\rm lock}\kappa^2_m}{\gamma\kappa_{mc}} =\frac{4}{3\sqrt{3}}\frac{\delta\omega_{\rm lock}}{\Gamma_m} ,
\label{kappa_do}
\end{equation}
where $\Gamma_m = 4\frac{\gamma\kappa_{\rm mc}}{\kappa_{\rm m}^2}$ is the amplitude reflection coefficient from the WGM cavity at resonance \cite{Kondratiev:17}, which can be measured experimentally. On the other hand, $\bar{\kappa}_{do} = \frac{T_o^2}{\tau_d R_e R_o^2}\sqrt{1+\alpha_g^2}$, where $R_o, R_e$ are the amplitude reflection coefficients of the output and end laser mirrors, $T_o = \sqrt{1-R_o^2}$ is the output amplitude transmission coefficient, $\tau_d$ is the round-trip time of a laser diode, $\alpha_g$ is the Henry factor, which allows taking into account the refractive index changes due to the injected carriers. So, the determination of the $\bar{\kappa}_{do}$ provides vital information about the laser diode itself. That way of $\bar{\kappa}_{do}$ estimation can be used not only for bulk resonators but also for on-chip ones.

\section{Near-IR experiment}
To verify the proposed technique, we performed some measurements of the FLR and BLR dependencies on the resonator-to-coupler gap (with the control of the Q-factor at zero-gap to determine the vertical index of the mode) for the MgF$_2$ microresonator at 1550 nm and compared the calculated quality factor with the result obtained with the well-known ringdown method.

\begin{figure}[htbp!]
\centering
\includegraphics[width=\linewidth]{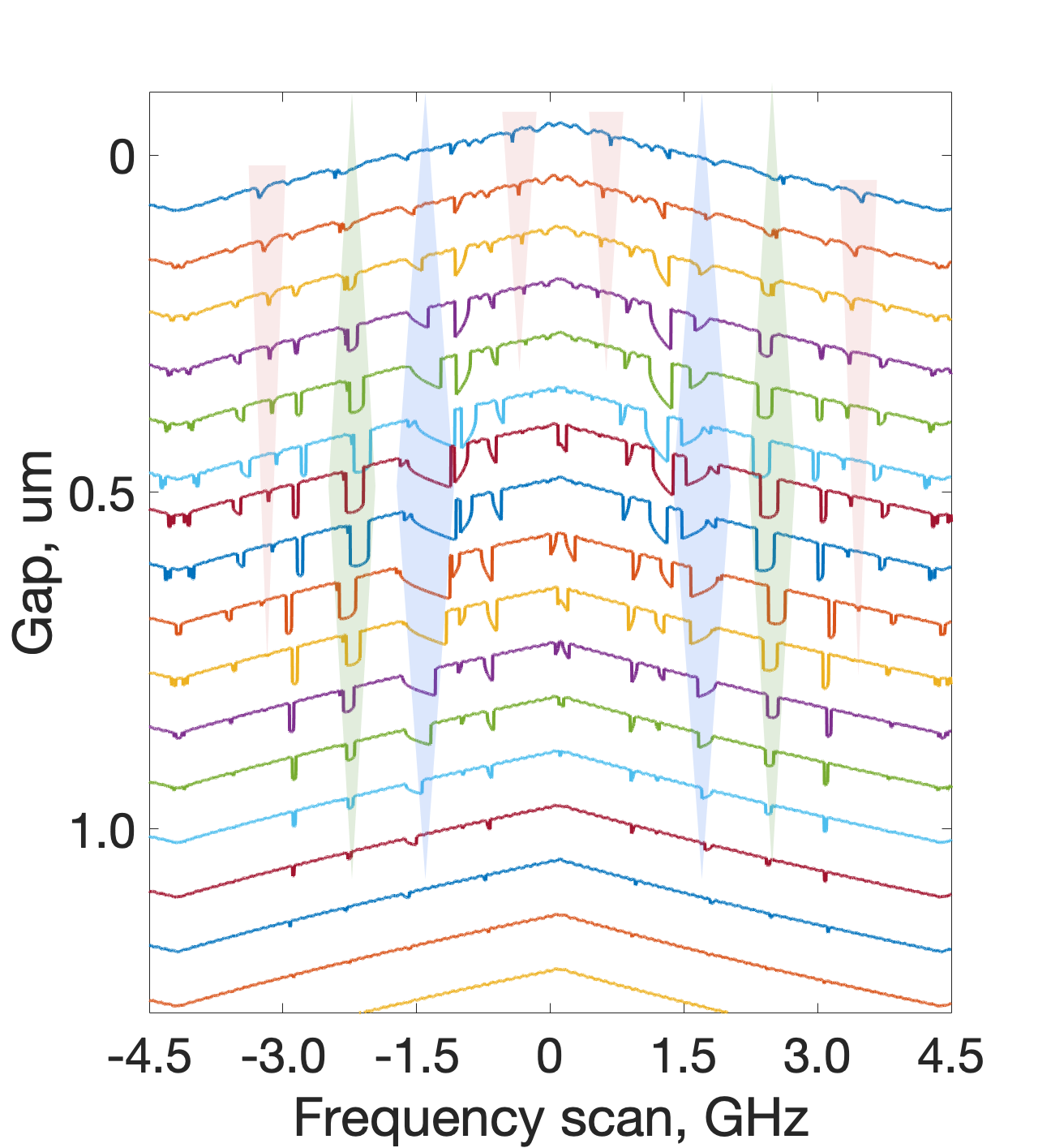}

\caption{LI-curves measured for different gap values. The highlighted regions corresponds to the excited modes with different polar indices: light green - $p = 0$, light blue - $p = 2$, light red - high $p$. The mode shapes remain constant upon variation of the gap value, which indicates that the locking phase $\psi_0$ stays unchanged. }
\label{fig:Modes}
\end{figure}

The WGMs were excited by a BK7 coupling prism with a DFB laser diode, exact wavelength was 1549.5 nm and power was below 2 mW. The microresonator was 4 mm in diameter and with 0.5 mm side curvature radius. The experimental setups and algorithm are presented in Fig. \eqref{fig:theBegining}.

\begin{figure*}[htbp!]
\centering
\includegraphics[width=\linewidth]{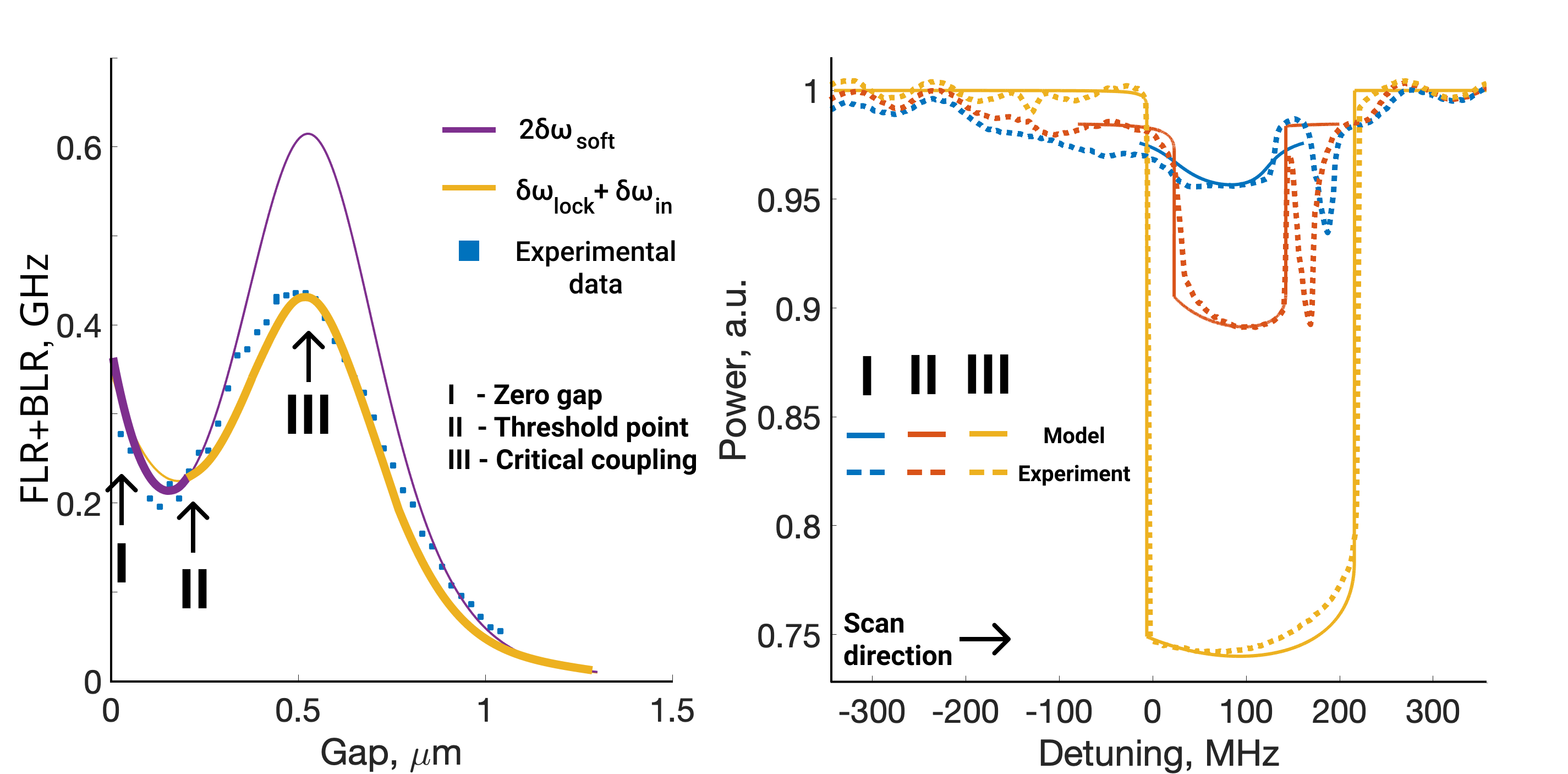}

\caption{Left panel: FLR+BLR dependence for the mode with $p = 0$. The transition to the "soft" regime occurs at a gap of around 0.2 um, and the fit with $\delta\omega_{\rm FWHM}$ is valid for gap values less than 0.2 $\mu$m. Right panel: the mode traces at three distinctive points: I - at zero gap, II - at transition point ($\rm{FLR}+\rm{BLR} = 4\kappa_m$) and III - at critical coupling point ($\kappa_{\rm mi}=\kappa_{\rm mc}$). The experimental curves are presented with dotted lines and the approximation curves made with the linear SIL model are solid lines. The high-order mode appeared at overcoupling. }
\label{fig:zerop}
\end{figure*}

\begin{figure*}[hbpt!]
\centering
\includegraphics[width=\linewidth]{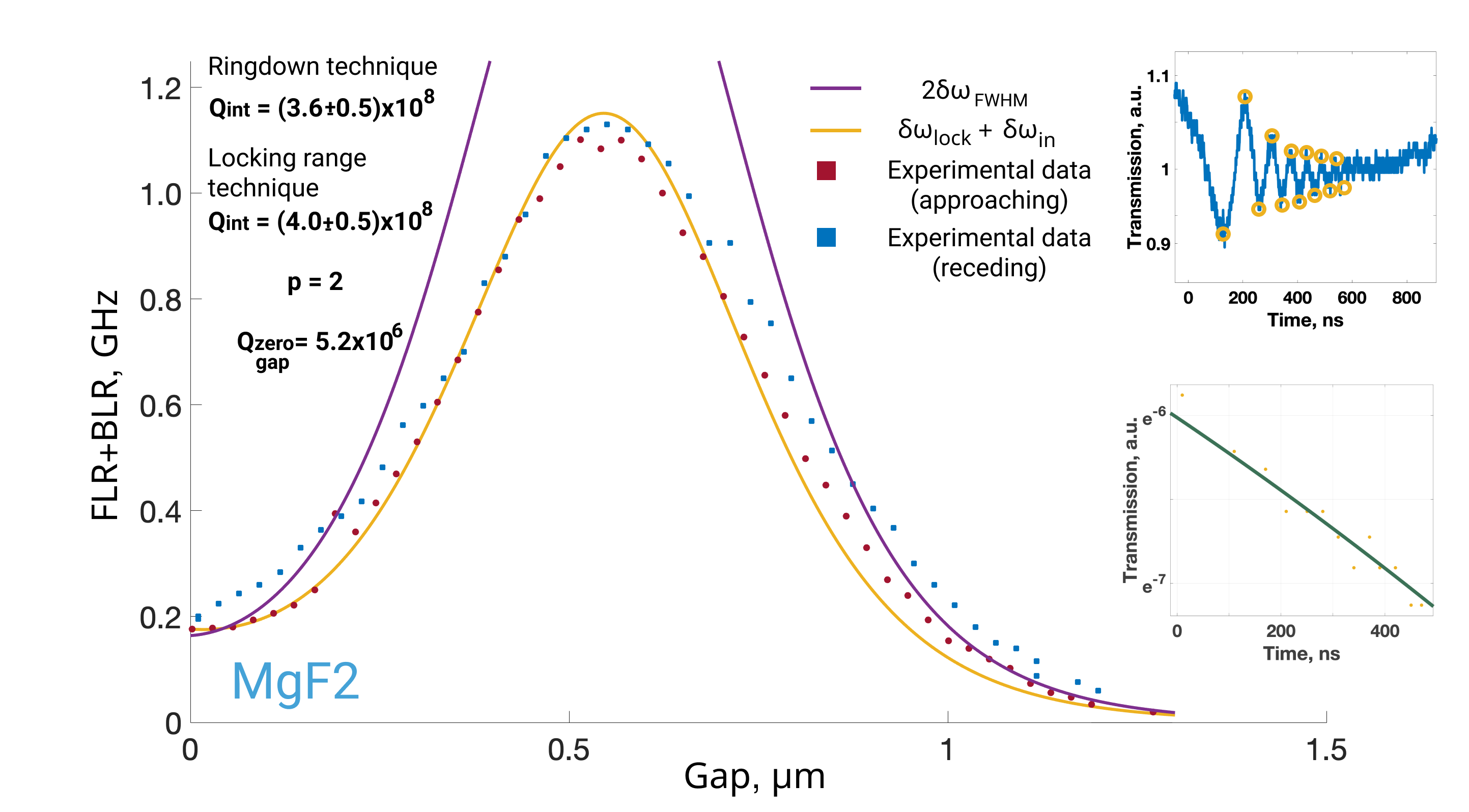}

\caption{Dependence of FLR+BLR on the gap for the $\rm{MgF_2}$ microresonator. The dots are the experimental data, the red ones are for approaching, and the blue ones are for receding microresonator with respect to the coupler. The yellow solid line is the $\delta\omega_{\rm{ lock}}+\delta\omega_{\rm{in}}$ curve calculated from \eqref{Lock} and \eqref{Inner}. The violet solid line is $2\delta\omega_{\rm {FWHM}}$ calculated from \eqref{SILFWHM}. The extremum takes place at 550 $\mu m$, which gives us $Q_{\rm int} = (4.0\pm0.5)\cdot10^8$ and $p = 2$. The ringdown method provides $Q_{\rm int} = (3.6\pm 0.5) \cdot10^8$, see the insets.  }
\label{fig:expCoupl}
\end{figure*}

The translation stage with piezo element (PZT) was previously experimentally calibrated, and the measured displacement coefficient was 0.27 $\pm$ 0.01 $\mu$m/V. The step size of the PZT was 27 nm with our controller. We measured the dependence of the locking width on the piezo voltage (which provided us the gap value) in both cases when frequency increased and decreased. FLR+BLR was measured using a calibration interferometer with 93 MHz free spectral range. The calibration interferometer was made with a 90:10 beamsplitter and mirrors. Observing the interference fringes, one may determine the frequency scale. The interferometer was continuously available, but during the measurements it was blocked to make the mode spectra clearer. For the locking width measurements we used a detector with large sensitive area 19.6 $mm^2$ and 460 kHz bandwidth, and for the ringdown measurements the detector with 0.8 $mm^2$ area and 700 MHz bandwidth. 
It is important that the locking phase doesn't change during the loading (see Fig. \eqref{fig:Modes}, where the transmission resonance curves are presented). Every line is like an "isogap", whose value one may find on the y-axis. In the mode traces, the modes with different $p$ and $Q$ are distinguishable. The modes from different WGM families were excited, so the different locking phases were observed on a trace. Colored areas in Fig. \eqref{fig:Modes} correspond to different types of the excited modes.   
The modes highlighted with red had high $p$ and lower $Q$ so that they appeared closer to the coupler and remained narrow. 
The mode highlighted with green had $p = 0$ and $Q_{\rm{int}} = 0.8\cdot10^8$, and its FLR+BLR dependence from the gap is shown in the left panel of Fig. \ref{fig:zerop}. A critical coupling point (maximum) and transition point ("hard" to "soft" SIL) are highlighted with arrows III and II correspondingly. There is a pronounced minimum in this case. It appears between the points where the FLR+BLR is wide due to the self-injection locking and overcoupled regime of the microresonator. At this point LI curve becomes narrower transforming to the "soft" regime while $\kappa_c$ is not large yet. We also have to note, that while $\delta\omega_{\rm lock}>\delta\omega_{\rm FWHM}$ in the "hard" regime, $\delta\omega_{\rm lock}+\delta\omega_{\rm in}<2\delta\omega_{\rm FWHM}$ and vice versa for the soft regime. We compare the LI curves at the three mentioned points with a linear SIL model approximation \cite{Kondratiev:17} in the right panel of Fig. \ref{fig:zerop}. 
First, we subtracted the transmission inclination due to the laser diode power change during the frequency scan from experimental data and normalized the signal to the off-resonant power. We note, that the experimental transmission in the critical coupling regime is 74\% (see dashed lines in the right panel of Fig. \ref{fig:zerop}) instead of close to zero value. This is due to imperfect mode matching in the experiment and the theoretical curves should be modified accordingly. The residual sinusoidal transmission oscillations refer to the calibration interferometer.
Then we calculated $Q_m$ and $Q_{\rm int}$ for the mentioned points and experimentally measured $\bar{\kappa}_{do}$. 
Using this data as input parameters for the modeling we carried out an approximation (see solid lines in the right panel of Fig. \ref{fig:zerop}) and get excellent agreement of the resonance curve form. The locking phase was initially assumed to be zero as forward and backward scans look symmetric (slight shift to $\pi/10$ makes the coincidence even better). However, after the alignment of the mode III we had to shift further the positions of the theoretical curves for the modes I and II to the right by 65 and 50 MHz for the perfect matching. We attribute this to the thermal effects, not included into the model, that shift the mode as a whole the more, the deeper is the dip (corresponding to the higher intracavity power). We also note that the power in the wide region near the mode drops down slightly. That probably happens due to the power leakage into some low-Q mode, revealed with loading. To account for this, we also lower the unity level for the theoretical curves I and II for 0.015 and 0.02. 
A mode from another family becomes visible in the curves I and II to the right from the locked one during the coupling increasing, while initially in III it was hidden by the locking. 
It was also revealed that the mode shape near the coupler depends on the $\bar{\kappa}_{\rm do}$: the lower it is, the closer the mode shape to Lorentzian.

In many cases the transmission to "soft" regime (without sharp edges) appeared, when the resonator was close enough to the coupler. It allows the direct determination of $p$ by the linewidth at zero gap (see Table \eqref{tab:shape-functions}). But in some cases, when the dimensionless coupling rate between counterpropagating modes and vertical index of the mode are high enough, the SIL regime can be observed at zero gap. After microresonator reached the surface of the coupler the resonance stays constant.

The mode highlighted in blue in Fig. \ref{fig:Modes} had a locking phase near $\pi$. We changed its the phase to zero by moving the laser and performed the measurements of its FLR+BLR dependence on the gap value, obtaining Fig. \eqref{fig:expCoupl}. Here the locking range is much wider and a transition from the "hard" to the "soft" regime is not easily noticeable. The extremum is at $d = 550$ $\mu$m from the coupler, which corresponds to an internal Q-factor $Q_{\rm{int}} = (4.0\pm0.5)\cdot10^8$ for $p = 2$. One may see that FLR+BLR is a bit wider when the microresonator recedes from the coupler, which is caused by the hysteresis of the piezo element. It is worth noting that there is a strong dependence of the extremum location on $p$, when it changes from $p = 0$ to $p > 0$. This dependence is weak for the high $p$ values, and for the $q$ index also \cite{gorodetsky1994high}. The transition point to "hard" locking regime for the modes with $p = 0$ is reached earlier than for $p > 0$, as $\kappa_c$ increases faster, the combined coupling coefficient decreases to values not enough for the strong SIL.

\begin{figure*}[hptb!]
\centering
\includegraphics[width=\linewidth]{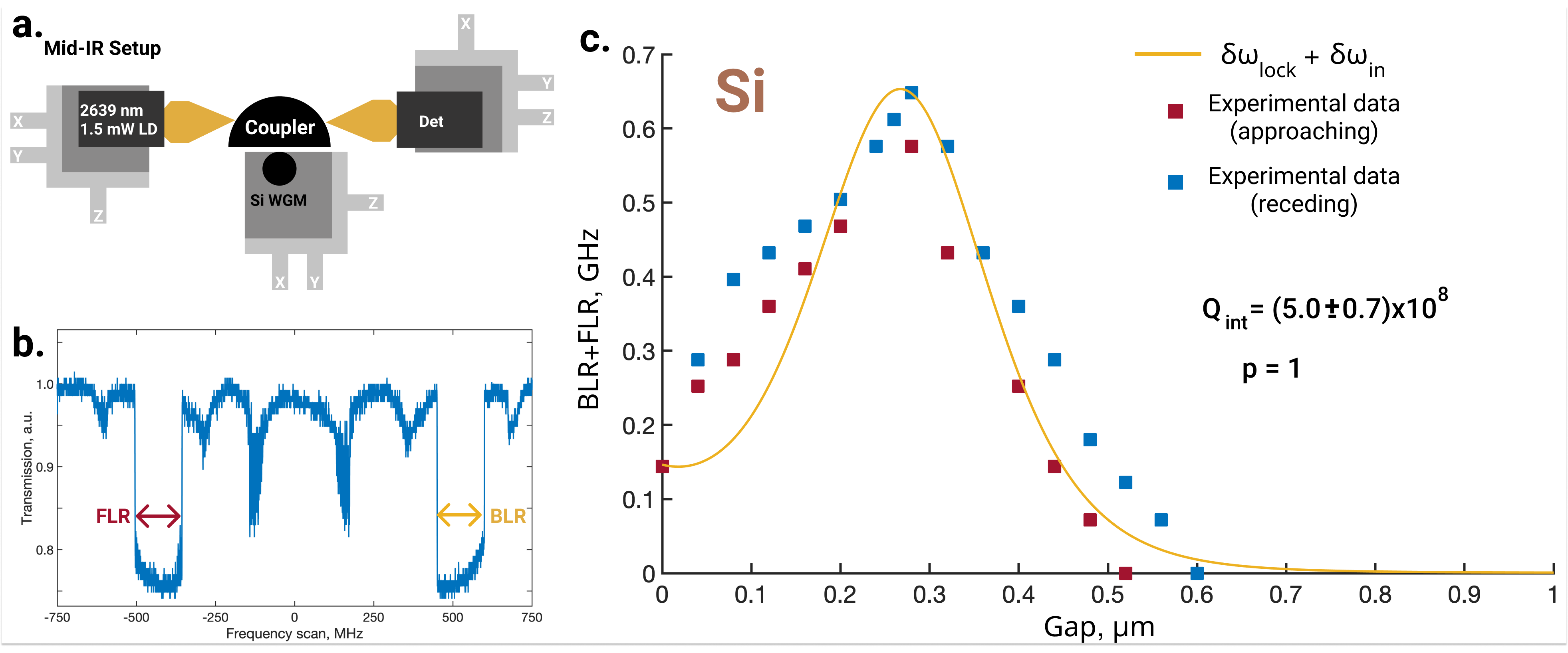}
\caption{(a) Experimental setup in the mid-IR. (b) Mode scan at the critical coupling. (c) Dependence of FLR+BLR on the gap for the Si microresonator at 2639 nm. The blue and red dots are the experimental data and the yellow solid line is a curve calculated using Eq. \eqref{Inner}. The extremum takes place at 270 $\mu m$ that gives us $Q_{\rm int} = 5.0\cdot10^8$ and $p = 1$. }
\label{fig:LRCoupling}
\end{figure*}

Finally, we performed quality-factor measurements of the mode highlighted in blue in Fig. \ref{fig:Modes} via the ringdown method. We used an external narrow linewidth fiber laser with an optical isolator. To excite the same WGM, we determine its exact frequency by beating transmitted light with laser with isolator at undercoupled regime. Then, we fix the polarization with a Glan-Thompson prism. Finally, we adjust the  power of the external laser with an isolator to make it equal to the DFB laser power. Thus, the mode was excited with the external laser with an isolator at the same frequency, the same polarization and at the same power. The WGM's intrinsic quality factor was measured by the ringdown method in the undercoupled regime as $Q_{\rm{int}} = (3.6\pm 0.5)\cdot 10^8$. To verify that there was no degradation of the $Q_{\rm{int}}$ during the measurements, we additionally checked that the locking width did not change after these measurements.

To measure the backscattering we added a 50:50 beamsplitter between the coupling lens and the external laser with an isolator. Half of the backscattering was coupled into the photodetector so we measured $\Gamma_m^2$. From the previous experiments we may evaluate the $\gamma\bar{\kappa}_{do}$ as a fit parameter. The backscattering wave measurements allows us to measure $\gamma$ as a part of $\Gamma_m$ separately from $\bar{\kappa}_{do}$. The measured reflected power for the critical coupling was 1/150 part of the input power ($\Gamma_m^2=1/150$). One may calculate $\bar{\kappa}_{d_o}/(2\pi) = 1.2\cdot10^{10}$ Hz.

The Q-factor measured with the proposed technique is in good agreement with the value measured by a conventional ringdown method, which confirms the applicability of the technique. The proposed technique provides a unique opportunity to accurately measure the Q-factor via SIL without changing the setup or introducing additional equipment.

\section{Mid-IR experiment}
The developed technique was applied to determine the Q-factor of a silicon microresonator at 2639 nm. The microresonator made of crystalline silicon was polished according to technique described in \cite{Shitikov:18} to provide low surface losses. First, we determined the internal Q-factor of this microresonator at 1550 nm as $1 - 2\cdot10^8$ depending on the mode. The Q-factor was measured using an external laser with an isolator by measuring the FWHM.

In the mid-IR experiment light was coupled into microresonator by a hemisphere made of silicon. The temperature-stabilized 2639 nm DFB diode laser was used as a pump laser. A temperature stabilized InAs photodiode with the amplifier with 10 MHz bandwidth and 100 $\mu$W saturation power was used as a detector. The immersive lens allowed to couple mm-scale beam into detector. Measurements of the WGM's Q-factor by the ringdown method are impossible here due to absence of the detector with high bandwidth, and limitations of the laser sweep speed. 
We conducted the experiments with an attenuating plate between the laser and a coupler to suppress the backscattering (and SIL) to determine the Q-factor by measuring the FWHM. Accurate  measurements of the quality factor were impossible due to the thermo-optical oscillations and laser noise. However, we estimated the linewidth of the microresonator at zero gap to be approximately $15$ MHz, which corresponds to $p > 0$. The proposed technique provides us with a unique opportunity to determine the microresonator's parameters via self-injection locking. The thermo-optical oscillations were drastically suppressed in SIL regime (see Fig. \eqref{fig:LRCoupling} (b)). The  internal Q-factor value of the WGM microresonator was as high as $5.0\pm0.7\cdot10^8$ and the vertical index of the mode $p = 1$ (see Fig. \eqref{fig:LRCoupling} (c)). 

At the zero gap there was a mode with sharp edges having width of about 70 MHz. The increase of the Q-factor by several times compared to the near-IR indicates that the main mechanism of the losses is the Rayleigh scattering. This confirms the assumption of attainability of ultra-high Q-factors in silicon WGM microresonators in the mid-IR. The effect of a wider measured locking range during taking microresonator from the coupler caused by hysteresis of the piezo element was also observed. The total locking range for small $d$ was higher than theory predictions. We suppose that this is due to the thermal effects in silicon, which are so noticeable because the coupler and microresonator are  both made of silicon, so that the slightest changes of the refractive index are marked. This effect was reduced, but did not vanish with the decrease of the incident power and for other modes. The total locking range exceeded $0.6$ GHz, that for $Q_{\rm{int}}= (5.0\pm0.7)\cdot10^8$ corresponds to the stabilization coefficient of $K = \frac{16\delta\omega_{\rm{lock}}}{3\sqrt{3}\kappa_m} = 4200$.

\section{Conclusion}
We developed, verified and implemented an original technique in an experiment that allowed us to determine the parameters of the reference microcavity in the SIL regime. The method is applicable for various spectral ranges, especially where it would be difficult to use established methods. The determination of the intrinsic Q-factor is achieved by measuring the locking width evolution upon changing the gap between the coupler and microresonator. In contrast to the method  based on the resonance depth analysis \cite{Coupling18} this technique allows for the Q-factor determination in the SIL regime, vertical mode index $p$ identification, and is not susceptible to stray light entering the photo detector. The information on the $p$ index of the mode increases the accuracy of the measurements. We showed experimentally that the phase of the SIL stays constant during the variation of the gap between the microresonator and the coupler. Measuring additionally the backscattering, we determined the effective output beam coupling rate of the laser, that is a key parameter for modeling a SIL laser.  

The technique was also used to determine the WGMs' parameters of silicon microcavity at 2639 nm, that was the first demonstration of the SIL in a silicon microcavity in the mid-IR. The elaborated approach allowed us to determine the internal Q-factor of the locking mode that was equal to $(5 \pm 0.7) \cdot 10^8$. Such a high value of the quality-factor directly observed in silicon verifies the significant possibilities of silicon microresonators for mid-IR photonics.

This work was supported by the Russian Science Foundation (project 20-12-00344).
AES and VEL acknowledge the Foundation for the Advancement
of Theoretical Physics and Mathematics “BASIS” for personal support.

\eject

\bibliography{SiSIL}
\end{document}